\documentclass[doublespacing]{elsart}

\usepackage{amsmath}
\usepackage{graphicx}

\begin{document}

\date{\today}

\begin{frontmatter}

\title
{Thermal effects in exciton harvesting in biased one-dimensional
systems}

\author{
S. M. Vlaming, V. A. Malyshev, J. Knoester}

\address{ Centre for Theoretical Physics and Zernike Institute for Advanced
Materials, University of Groningen, Nijenborgh 4, 9747 AG Groningen,
The Netherlands }


\corauth[corauthor]{ Corresponding author: J. Knoester, Centre for
Theoretical Physics and Zernike Institute for Advanced Materials,
University of Groningen, Nijenborgh 4, 9747 AG Groningen, The
Netherlands; tel: +31-50-3634369; fax: +31-50-3634947; email:
j.knoester@rug.nl }

\begin{abstract}

The study of energy harvesting in chain-like structures is important
due to its relevance to a variety of interesting physical systems.
Harvesting is understood as the combination of exciton transport
through intra-band exciton relaxation (via scattering on phonon
modes) and subsequent quenching by a trap. Previously, we have shown
that in the low temperature limit different harvesting scenarios as
a function of the applied bias strength (slope of the energy
gradient towards the trap) are possible~\cite{Vlaming07}. This paper
generalizes the results for both homogeneous and disordered chains
to nonzero temperatures. We show that thermal effects are
appreciable only for low bias strengths, particularly so in
disordered systems, and lead to faster harvesting.

\end{abstract}


\journal{J. Luminescence}
\end{frontmatter}

\newpage

\section{Introduction}

The study of energy transport and trapping in molecular scale
systems is an interesting topic, because of its relevance to natural
systems (photosynthetic antenna complexes \cite{Mukai99}) and
applications in nanophysics and quantum computing where energy
should efficiently be transported over some distance. Linear
molecular aggregates~\cite{Kobayashi96,Knoester02}, conjugated
polymers~\cite{Tilgner92,Hadzii99}, photonic
wires~\cite{Wagner94,Heilemann04} and quantum dot
arrays~\cite{Crooker02,Franzl04} may all be envisioned as
realizations of such energy transport complexes. A natural question
that arises is how this combined process of transport and eventual
trapping, which we will from now on refer to as harvesting, can be
optimized. In a previous paper~\cite{Vlaming07}, we performed a
study on the effect of an energy gradient (bias) on the energy
harvesting properties of a linear system. The scope there was
limited to low temperatures. In this paper, we present numerical
calculations of thermal effects on the harvesting time and a
qualitative analysis of the results.

\section{Model}

We use a Frenkel exciton model to describe the excited state
dynamics in chain-like systems. The latter are modeled as a regular
one-dimensional array of two-level monomers that are coupled through
their transition dipole moments, which are assumed to be oriented
parallel to each other. We initially excite a monomer at one end of
the chain ($n=N$), and consider the other end ($n=1$) as a trap. The
harvesting time is calculated as the average time it takes for an
excitation to move across the chain and to be quenched by the trap.
Including diagonal disorder and a bias ($\Delta>0$) towards the
trap, the model Hamiltonian reads
\begin{equation}
    \label{frenkelham}
    H_{ex}=\sum_{n=1}^N [\varepsilon_n + (n-1)\Delta]\left|n\right>\left<n\right|
    + \sum_{n,m\not=n}^N J_{nm} \left|n\right>\left<m\right|,
\end{equation}
where the state $\left|n\right>$ denotes that monomer $n$ is excited
while all other monomers are not, the excitation energies
$\varepsilon_n$ are uncorrelated Gaussian stochastic variables with
zero mean and standard deviation $\sigma$ and the transfer integrals
are $J_{nm}=-J/|n-m|^3$, with the nearest-neighbor coupling $J > 0$.
Diagonalization of this Hamiltonian yields collective excited
states, $\left|s\right>=\sum_n c_{sn}\left|n\right>$. Both a bias
and disorder result in localization of the states, Bloch-like
\cite{Bloch28} and Anderson-like \cite{Abrahams79}, respectively, so
that the exciton states are spread out over finite segments of the
chain only. In the case of a bias, there exists a correlation
between the location of the eigenstate and its energy, as lower
energy states are located in the vicinity of the trap.

The dynamics of the populations $P_s$ of the various exciton states
is analyzed within the framework of the Pauli master equation,
\begin{equation}
\label{PME}
    \dot{P}_s = -(\gamma_s + \Gamma_s)P_s
    + \sum_{s'}\left( W_{ss^{\prime}}P_{s^{\prime}}
    - W_{s^{\prime}s}P_s \right) \ ,
\end{equation}
where $\gamma_s=\gamma_0\left(\sum_n c_{sn}\right)^2$ is the
radiative rate of the state $\left|s\right>$ ($\gamma_0$ being the
radiative rate of a monomer) and $\Gamma_s$ is the quenching rate,
which is taken proportional to the exciton probability at the trap
($n=1$):~$ \label{quench1} \Gamma_s = \Gamma |c_{s1}|^2 $ where
$\Gamma$ is the quenching amplitude~\cite{Malyshev03}, which can not
be arbitrarily large within the limitations of our
model~\cite{Vlaming07}. As at $t=0$ only the rightmost site $n=N$ is
excited, the initial population distribution is given by
$P_s(0)=|c_{sN}|^2$. The $W_{ss'}$ describe the scattering of
population between various exciton states through interaction with
phonons:~$ W_{ss^{\prime}} = W_0\ S(E_s - E_{s^{\prime}})
\,\sum_{n=1}^N c_{s n}^2 c_{s^{\prime} n}^2 G(E_s-E_{s'}), $ where
$W_0$ is the scattering amplitude, $S(E)$ is the phonon spectral
density function, $\sum_{n=1}^N c_{s n}^2 c_{s^{\prime} n}^2$ is a
probability overlap, and $G(E)=1+n(E)$ if $E<0$, while $G(E)=n(E)$
if $E>0$ with $n(E)=\left[\exp(|E|/T)-1\right]^{-1}$ the phonon
occupation factor. We use a Debye-like spectral density with cut-off
$\omega_c$, $S(E)=|E/J|^3\exp(-|E|/\omega_c)$. The harvesting time
is defined as follows:
$\tau^{-1}=\tau^{-1}_{\mathrm{trap}}-\tau_0^{-1}$, where
$\tau_{\mathrm{trap}}$ and $\tau_0$ are the decay times with and
without trap, respectively. They are calculated as
$\int_0^{\infty}dt\left<\sum_s P_s(t)\right>,$ where
$\left<...\right>$ denotes the disorder average.

\section{Disorder-free biased chains}

We begin with a discussion of the harvesting efficiency in
homogeneous chains ($\sigma=0$). The bias-induced Bloch localization
dramatically changes the harvesting properties. Depending on the
relative strengths of the quenching and relaxation processes, both
monotonic decreasing and nonmonotonic behavior of the harvesting
time as a function of the bias strength can be obtained. We refer to
our earlier paper~\cite{Vlaming07} for a full discussion of the
possible scenarios, while here we focus on the aspects relevant to
understanding thermal effects in our model.

\begin{figure}[ht]
\centerline{\includegraphics[width=8cm]{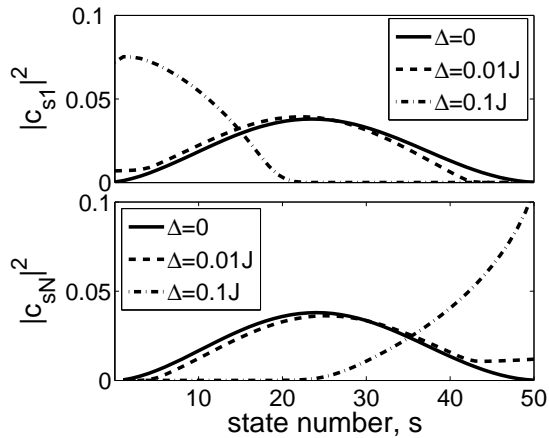}}
    \caption{Bias dependence of the exciton probabilities $c_{sN}^2$
    and $c_{s1}^2$ that determine the initial distribution of the
    exciton population $P_s(0)$ and the quenching rate $\Gamma_s$,
    respectively, for chains without disorder ($\sigma=0$). The
    bias tends to shift the initial distribution
    of population to the top of the band (to higher $s$). Oppositely,
    the strongly quenched states occur at the lower band edge (for small
    $s$). }
\label{fig:f3}
\end{figure}

\begin{figure}[ht]
\centerline{\includegraphics[width=10cm]{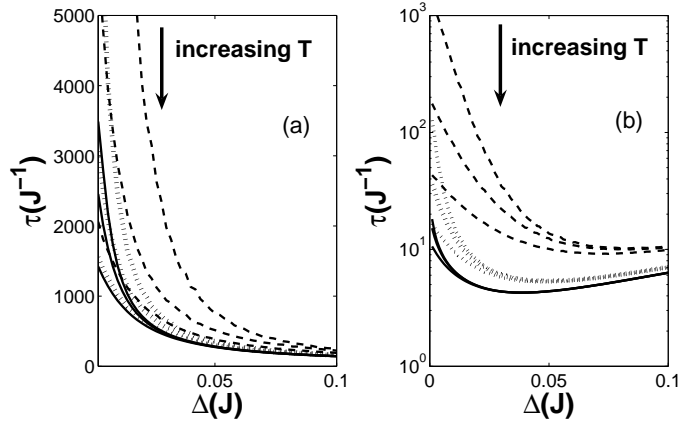}}
    \caption{Harvesting time $\tau$
    versus the applied bias magnitude $\Delta$ calculated for
    disordered chains of 50 sites for the quenching amplitude
    $\Gamma=10^{-1}J$ (panel (a)) and $\Gamma=10J$ (panel (b)) and the scattering amplitude $W_0 = 10
    J$. The curves were calculated for different temperatures,
    $T=0.02J$, $T=0.2J$ and $T=0.4J$, which are ordered as shown by
    the arrow. Solid, dotted and dashed curves correspond to
    disorder strength $\sigma=0$,~$\sigma=0.2J$ and $\sigma=0.4J$,
    respectively.
} \label{fig:quench}
\end{figure}

In the limit of $\Gamma \ll W_0$ and for low temperature, relaxation
to lower exciton states is the dominant process for virtually all
exciton states. As a result, most population ends up in the bottom
states of the band. For zero bias, the quenching rate from these
states is low because of the poor overlap of the bottom states with
the trap (see Fig. \ref{fig:f3}). However, turning on the bias
increases this overlap, thereby accelerating the harvesting process.
Temperature will enhance the harvesting efficiency at low bias by
shifting population from the poorly quenched bottom state(s) to more
strongly quenched higher energy states. This enhancement disappears
for higher bias strengths, as the higher energy states are no longer
more strongly quenched than the bottom states, as is shown in Fig.
\ref{fig:quench}(a).

The interesting situation where $\Gamma \sim W_0$ provides a
nonmonotonic bias dependency of $\tau$. A small bias is advantageous
for the harvesting process as it increases the quenching rate for
the bottom states, but beyond some optimal bias value the increase
in relaxation time overtakes the previous effect in importance and
the net effect of increasing the bias becomes negative. The
thermally induced changes in the harvesting efficiency are again
only present for small values of the bias, for the reasons mentioned
in the discussion of the limit of $\Gamma \ll W_0$. This is
corroborated in Fig. \ref{fig:quench}(b).

In the limit $\Gamma \gg W_0$ our model does not adequately describe
the harvesting, in particular when the exciton states are
delocalized over the entire chain (i.e., small bias, small disorder,
and relatively short chains). In this situation, the off-diagonal
density matrix elements (exciton coherences), which are also created
by populating the site $n=N$ [$\rho_{ss'}(0)=c_{sN}c_{s'N}$], are
sufficiently long-lived to be relevant to the transport process. In
fact, neglecting coherences leads to the unphysical artifact that
there is instantaneous energy transfer over the chain. A more
refined approach is necessary to remove the occurrence of
unphysically fast energy transfer; we refer to our earlier
paper~\cite{Vlaming07} for a more detailed discussion of this limit
and its complications.

\section{Disordered biased chains}

The inclusion of disorder slows down the energy transport process
for all bias values; however, the effect is especially strong for
small biases. This will lead to a shift of the optimal bias (when
$W_0 \sim \Gamma$) to higher values with increasing disorder
strength. The overall shape of the $\tau(\Delta)$ curve does not
change in most cases: only in the case of a high quenching amplitude
Fig. \ref{fig:quench}(b), we see that an optimal bias appears as a
result of the fact that the disorder does more damage for low biases
than for higher ones.

The explanation of these effects is straightforward. Disorder
results in Anderson localization, even in the absence of a bias.
This leads to reduced relaxation rates, since the overlap of
different states decreases. The effect is particularly significant
for systems with a low bias, since in a disorder-free chain the
excitons are then delocalized over the whole chain. For higher
biases, there is already considerable localization because of the
bias itself, and the effect produced by the disorder is relatively
small. More importantly, for sufficiently strong disorder and small
bias the lowest energy states are not necessarily located near the
trap. At low temperatures, a large population is built up in these
states and the excitation may thus be unable to be trapped
efficiently because of the blockage of diffusion
\cite{Malyshev03,Bednarz03}. A bias forces these low energy states
to be located near the trap and thus accelerates the harvesting.

Thermal effects are marginal for large bias strengths. In this case,
the harvesting process consists largely of downward relaxation
followed by quenching from the bottom states. These states tend to
be localized near the trap and the quenching process is therefore
hardly accelerated by thermally induced population shifts to higher
energy states. In contrast, the thermal effects for small bias are
quite strong in disordered systems, even more so than in homogeneous
chains. As has been noted in the previous paragraphs, population
tends to end up in the exciton states at the bottom of the band.
These, in general, are not located near the trap and thus are
quenched poorly. Increasing the temperature has two effects. First,
it rapidly accelerates the exciton diffusion, and second, the
population is thermally excited to higher energy states, which are
quenched more strongly due to a better overlap with the trap. In
other words, the interaction of excitons with the phonon bath can
counteract some of the detrimental effects of Anderson localization
on the harvesting process, as can be seen in Fig. \ref{fig:quench}.

\section{Conclusions}

Contrary to what one might naively expect, namely that an energetic
bias towards the trap will decrease the harvesting time, we have
shown that various scenarios are possible, depending on the relative
strengths of the relevant subprocesses. This is already apparent in
low temperature systems (with and without
disorder)~\cite{Vlaming07}, and the same holds for systems at
elevated temperature. Thermal effects are only significant for small
bias strengths and tend to accelerate harvesting, especially in
disordered systems.

\end{document}